# Network Evolution of Body Plans

Koichi Fujimoto[1,2] [*], Shuji Ishihara[2,3] and Kunihiko Kaneko[1,2]

[1] ERATO Complex Systems Biology Project, Japan Science and Technology Agency;

[2] Department of Basic Science, University of Tokyo, 3-8-1 Komaba, Meguro, Tokyo 153-8902 Japan ;

[3] Division of Theoretical Biology, National Institute for Basic Biology, 5-1 Higashiyama, Okazaki, 444-8787, Japan.

[*] Corresponding author: fujimoto@complex.c.u-tokyo.ac.jp

**Abstract:**

One of the major goals in evolutionary developmental biology is to understand the relationship between gene regulatory networks and the diverse morphologies and their functionalities. Are the diversities solely triggered by random events, or are they inevitable outcomes of an interplay between evolving gene networks and natural selection? Segmentation in arthropod embryogenesis represents a well-known example of body plan diversity. Striped patterns of gene expression that lead to the future body segments appear simultaneously or sequentially in long and short germ-band development, respectively. Moreover, a combination of both is found in intermediate germ-band development. Regulatory genes relevant for stripe formation are evolutionarily conserved among arthropods, therefore the differences in the observed traits are thought to have originated from how the genes are wired. To reveal the basic differences in the network structure, we have numerically evolved hundreds of gene regulatory networks that produce striped patterns of gene expression. By analyzing the topologies of the generated networks, we show that the characteristics of stripe formation in long and short germ-band development are determined by Feed-Forward Loops (FFLs) and negative Feed-Back Loops (FBLs) respectively, and those of intermediate germ-band development are determined by the interconnections between FFL and negative FBL. Network architectures, gene expression patterns and knockout responses exhibited by the artificially evolved networks agree with those reported in the fly *Drosophila melanogaster* and the beetle *Tribolium castaneum*. For other arthropod species, principal network architectures that remain largely unknown **are predicted. Our results suggest that the emergence of the three modes of body segmentation in arthropods is an inherent property of the evolving networks.**

**Introduction**

Evolutionary diversification of multi-cellular organisms largely depends on body plans, in which complex morphologies develop under the integrated control of multiple genes [1]. The interaction among genes and gene products forms a regulatory network that orchestrates gene expression pattern to specify the morphologies. Mutational modification in gene regulation networks alters gene expression dynamics that provide a basis for morphogenetic diversity. A fundamental key to understanding evolutionary developmental biology is to elucidate how a gene network determines body plan, its diversity, and its potential to evolve [2-6]. Here we focus on gene expression patterning in segmented body plans during arthropod embryogenesis as model systems to address this question.

Arthropod segmentation exhibits three developmental modes of the stripe pattern formation in gene expression that specify the future elementary segments of an adult body [7,8]. Many of the descendant arthropod species (Fig. 1A; e.g., the fly *Drosophila melanogaster* [9]) follow the 'long germ-band' mode of development where stripes appear simultaneously along the anterior-posterior axis. In contrast, ancestral species (Fig. 1B; e.g., the beetle *Tribolium castaneum* [10] and the spider *Cupiennius salei* [11]) exhibit 'short germ-band' mode where stripes appear sequentially. A combination of both is found in 'intermediate germ-band' mode; anterior stripes appear simultaneously while the remaining posterior stripes appear sequentially (Fig. 1C; e.g., the cricket *Gryllus bimaculatus* [12] and the milkweed bug *Oncopeltus fasciatus* [13]). Conservation of regulatory genes such as gap and pair-rule genes among arthropods indicates that the differences in the stripe formation have originated from architecture of the regulatory network. Comparative studies from species to species have extensively been carried out to reveal differences in spatiotemporal gene expression pattern while knockout responses are studied to decipher a functional role of genes in shaping the morphogenesis [14-17].

These observations raise three related problems. First, what is basic difference in network architecture that distinguishes the three modes? Second, how does a distinct network architecture produce spatio-temporal gene expression corresponding to each developmental mode for segmentation? Can the functional role of each network architecture account for observed knockout responses? Third, what type of evolution pressure will favor the selection of each developmental mode? So far the understanding of the evolution of gene regulatory networks remain too fragmentary to answer these questions, due to practical limitations of time scale in experimental approaches.

To address these problems, here we adopt an integrated approach by analyzing structure and function of gene networks, and modeling diversity in striped pattern formation. In order to reveal the basic differences in the network architecture, developmental gene networks are numerically evolved [18-23] under selection pressure to form a target number of stripes expressed in a specific gene, which we label #1 without loss of generality (Figure S1; see Methods). We find emergence of three developmental modes to form the stripes. The three modes are characterized by the presence and abundance of Feed-Forward Loops (FFLs), Feed-Back Loops (FBLs), and interconnection between the two types of loops in the gene network. As we will see later, these three modes strikingly agree with long, short, and intermediate germ development in arthropod segmentation respectively, with regard to spatio-temporal gene expression and knockout responses. Furthermore, network architectures composed of FFLs and/or negative FBLs exhibit a trade-off constraint between mutational robustness and developmental speed, which may play a crucial role in the evolution of segmented body plans.

**Results and Discussions**

**Three developmental modes in artificial evolution.**

Within approximately 1000 independent evolutionary trials, we discovered that the selected networks exhibit three basic modes of spatio-temporal gene expression (Figs. 1D-F and S12): simultaneous, sequential, and combinatorial stripe formation. In the mode displayed in Figure 1D, stripes appear almost simultaneously, while in another mode shown in Figure 1E each stripe appears one by one. Figure 1F shows an example of combinatorial formation, where stripes appear simultaneously on the left side but sequentially on the right side. These modes are well known for the spatio-temporal expression of segment polarity genes in the long [9,24,25] (Fig. 1A), short [10,11,26-29] (Fig. 1B), and intermediate [12,13] (Fig. 1C) germ embryogenesis of arthropods. In addition to simultaneous stripe formation of gene #1, expression of the upstream genes in the network (Fig. 2A) also follows a characteristic pattern observed in long germ insects [9,24,25] (Figs. 1G and S2A); a maternal gene in a simple gradient, gap genes in one or two domains, pair-rule genes that form half as many stripes as segment polarity genes – a phenomenon known as 'double segment periodicity [9,16]'. Similarly, in networks exhibiting sequential and combinatorial stripe formation, as will be discussed, the expression patterns of the other genes closely follow those reported for short [30] and intermediate [12,13,31] germ-band arthropods respectively (Figs. 1H-I, S2B-C, and S3B-C).

**Modularity in artificially evolved networks.**

In order to find the underlying network properties that give rise to the three distinct developmental modes, we first extracted minimal sub-networks necessary for the striped pattern (Fig. 1D-F) from the evolved networks (Fig. 2A-C; see Methods and other representative examples in Fig. S5). We shall hereafter refer to these as 'core networks'. Second, the core networks were classified into long, short, and intermediate germ modes according to the exhibited mode of stripe formation as described above (see Methods). Then, for each mode, we investigated the appearances of the two prominent motifs in regulatory networks - FFLs and FBLs [32-41]. We have discovered that multiple FFLs (Fig. 2A) are always included in the core networks in the long germ modes while at least one negative FBL (Fig. 2B) is always included in the short germ mode. Figure 2D shows the fraction of core networks that contains FFL and negative and positive FBLs. Multiple occurrences of FFLs in the long germ network (indicated by green bar graph in Fig. 2D) have been observed, while the appearance of at least one negative FBL in the short germ network (indicated by pink in Fig. 2D; Positive FBL is not always included in either long or short germ networks as indicated by gray in Fig. 2D). Both FFL and negative FBL always coexist for the intermediate germ network (Fig. 2C-D).

**Mechanism of striped pattern formation based on FFLs and FBLs.**

A single FFL functions as a stripe generator [42-44] (see Supporting Result S1 for a theoretical analysis). Let us give an example by examining a FFL from gene #0 to #30 in Figure 2A. The FFL lies downstream of maternal factor #0 that is imposed in the form of a simple gradient. Since gene #30 is activated by gene #26 and at the same time repressed by gene #5 depending on the level of #0, expression of #30 appears in a single stripe (Figs. 1D and S2A). The function of FFLs connected in series (marked by * in Fig. 2A) is to double the number of stripes, whereas the function of FFLs connected in parallel (marked by + in Fig. 2A) is to add a stripe [42]. The number of stripes to be added is determined depending on the number of FFLs connected in series or in parallel (Figs. S7 and S8). A negative FBL, on the other hand, functions as a temporal oscillation generator. Short germ development is expected to operate by a mechanism [14,16] similar to segmentation in vertebrates where oscillations are translated into sequential striped patterns by intercellular interactions [45-49]. Genes located either within or directly downstream of a negative FBL are subjected to temporal regulation by the FBL (Fig. S9B), resulting in sequential stripe formation (Fig. S3B). In the intermediate germ mode, genes regulated by a negative FBL (marked by Δ in Fig. 2C) show the sequential stripe formation, whereas genes regulated by FFLs (marked by + in Fig. 2C) show simultaneous stripe formation (Figs. S3C and S9C). These results suggest that parallel connection of FFL and negative FBL organizes the combinatorial stripe formation.

We examined the roles of FFL and FBL by performing 'knockout experiments' in all evolved networks (see Methods). The stripes in gene #1 vanish by eliminating a gene or a connection either within or downstream of a FFL or FBL. Perturbations of a FFL connected in parallel (+ in Fig. 2A and C) often results in defects confined to a few domains in a long or intermediate germ mode as observed for the gap mutation [12,13,50] (yellow green panels in Fig. 3). Disrupting a FFL connected in series (* in Fig. 2A-C) often leads to absence of every other stripes as in the pair-rule mutation [30,50] (blue green panels in Fig. 3), while disrupting gene at the top of the FFL (e.g., #14 in Fig. 2A) extinguishes all the stripes (the lowest figure in Fig. 3A). By disrupting a negative FBL ($\Delta$ in Fig. 2B-C), stripes that are formed sequentially are extinguished completely in short and intermediate germ modes (pink panels in Fig. 3).

The function of positive FBL sharpens striped pattern through interaction with a FFL [51] and amplifies temporal oscillation through the interaction with a negative FBL. However each role of positive FBL can be substituted by FFL and negative FBL, respectively, by tuning up parameter values in the FFL and the negative FBL through evolution. Thus a positive FBL is not necessary module (Fig. 2D). These results indicate that FFL and negative FBL are elementary modules responsible for the three characteristic modes of development.

**Network architecture in arthropod segmentation.**

In contrast to detailed models for a specific species [52,53], our aim is to capture general consequence of evolution of gene expression dynamics that hold over a large number of both artificial and arthropod networks. All the evolved network models we examined were exactly classified into three modes, sequential, simultaneous or combinatorial formation, respectively. We identified necessary network module for each mode (Fig. 2D) and confirmed its function for the stripe formation (Fig. 3A-C and Result S1). Characteristics in spatiotemporal gene expression pattern and the network structure are summarized in Table 1. These three modes in our models agree rather well with the short, long, and intermediate modes in arthropods.

Strikingly, besides the above correspondence in segmentation modes, we almost always find genes that qualitatively agree with arthropod genes in terms of how, where and in what order these genes are being expressed (Figs. 1G-I, S2, and S3). Moreover, when these genes are deleted from the network and compared with the respective knockout mutants in real arthropods, the altered expression patterns of gene #1 (Fig. 3A-C) and the segment polarity genes exhibit remarkable similarities (Fig. 3D). By focusing on the function of FFL and negative FBL, where the networks modules are located in the arthropod gene regulatory networks and how the arthropod genes are wired are straightforwardly inferred from mapping them to the corresponding genes in the artificial networks.

**Gap genes:** As shown in Fig. 1G and I, several genes express in a few domains generated by FFL connected in parallel (see also Fig. S7A-C, and 2nd figure in Fig. S2A and C). Whenever one of the genes is disrupted, a defect of striped pattern is produced locally for a corresponding domain (yellow green panels in Fig. 3). For example, such response is shown in the knock-out of gene #30 and #17 in Fig. 3A, and #25 in Fig. 3C. Indeed, these types of expression pattern in wild type and local defect of stripes induced in segmentation gene are known as roles of gap genes in a long germ insect *D. melanogaster* [49], and a gap gene *Krüppel* in intermediate germ insects *G. bimaculatus* [12] and *O. fasciatus* [13]. Even though detailed knowledge on the gene network for them is not yet available, we infer here that the arthropod genes should be located within a FFL connected in parallel, as in #30 and 17 in Fig. 2A, and #25 in Fig. 2C.

**Pair-rule genes:** In our models, several genes exhibit the double segmental periodicity generated by FFL connected in series where the stripe number is as half as that of segmentation gene #1 (Fig. 1G-I). Disrupting one of the genes located within the FFL always leads to absence of every other stripe with deletion of odd- or even-numbered stripe (blue green panels in Fig. 3B) or fusion of each pair of two stripes (Fig. 3A and C), while disrupting a gene at the top of the FFL extinguishes the stripes (the lowest panel in Fig. 3A). For example, the former response appears by the knock-out of gene #27 in Fig. 3A, #20 in Fig. 3B and #14 in Fig. 3C, whereas the latter by the knockout of gene #14 in Fig. 3A. Both the double segment periodicity and the mutant phenotype emerge as a result of the FFL connected in series (Fig. S7D). Indeed, the double segment periodicity is widely observed in arthropod pair-rule gene expression [9,10,26,29,30,54] . Disrupting the secondary pair-rule genes [9] in *D. melanogaster* and *T. castaneum* (short germ) leads to absence of every other stripes in segment polarity gene expression with the deletion [30,50] or the fusion [50,55] of every other stripe, while null mutation of the primary pair-rule gene *even-skipped* in *D. melanogaster* extinguishes the segments [56]. Thus the arthropod secondary and primary pair-rule genes are expected within a FFL connected in series (e.g, #27 in Fig. 2A, #20 in Fig. 2B and #14 in Fig. 2C), and at the top of the FFL (#14 in Fig. 2A), respectively.

**Genes which express striped pattern sequentially:** In short germ network models, several genes in a negative FBL express striped pattern sequentially from the anterior to posterior end while disrupting one of the genes always extinguishes almost all the stripes (e.g., gene #10, #11, #13 and #17 in Figs. S3B and 3B). In intermediate germ network models, a gene subjected to a negative FBL expresses striped pattern sequentially around posterior end while disrupting the gene extinguishes the stripes at the corresponding domain in the wild type (#3 in Figs. S3C and 3C). Moreover, striped pattern among genes in the FBL is partially overlapped, irrespective of the developmental modes (e.g., Fig. S2B). We have found such partial overlap only when the genes are located in a negative FBL (Δ in Fig. 2B). Indeed, these types of spatio-temporal expression and knockout responses were reported in primary pair-rule genes in *T. castaneum* [30,57], *Notch/Delta* in *C. salei* (short germ) [11], and *even-skipped* in *O. fasciatus* [31]. Thus these arthropod genes are expected to be located either within (e.g., #10, #11, #13 and #17 in Fig. 2B) or at the downstream of a FBL (#3 in Fig. 2C).

Abundance and interconnection of FFLs in accordance with the above predictions are well documented in *D. melanogaster* [42,58]. For example, existence of FFL composed of primary and secondary pair-rule genes and segment polarity gene was reported (Fig. 5 in ref. [55]). For *T. castaneum* [30], genetic studies suggest that the primary and secondary pair-rule genes are located within a negative FBL and a FFL connected in series as shown in Figure 2B. We infer that the negative FBL and FFL are responsible modules for forming stripes sequentially and double segmental periodicity, respectively. Spatio-temporal expression and knockout response of evolutionarily conserved genes such as *even-skipped* may differ dramatically from species to species [8,16,17,59]. The above results exemplify the necessary rewiring of FFLs and/or negative FBLs that must have taken place during the arthropod evolution.

**Network modularity and the robustness in developmental evolution.**
We now discuss implications of the network architectures derived from our models to each developmental mode and evolutionary process. The hierarchical structure of FFLs add or double stripes in order to form multiple stripes in all long germ core networks; a gene expressed in a simple gradient (#10 and #26 in Figs. 1G, S2A and 2A, and Result S1) is followed by genes that are expressed in one or two stripes (#30 and #6). They are further connected to genes appearing in many more stripes (#14 and #1). The knockout response varies depending on the exact position of the disturbed FFL in the core network (Fig. 3A). On the other hand, variations in striped pattern are only occasionally observed in short germ networks. The majority of the mutant networks show no changes in the number of stripes while a very small fraction of them fails to form stripes all together (Fig. 3B). Hence, a hierarchy of FFLs and a variety of knockout responses are necessary features of the long germ development. In contrast, for the short term development, there is no such hierarchy and consequently, no strict necessity in variety of knockout response.

The susceptibility to network perturbation (Fig. 3) is known as robustness of the network [21-23,52,60-63]. The small size of the core network (Figs. 4A and S6C-D) implies less chance for the dynamics to be disrupted by mutation. Of course how a certain gene regulatory network works depends not only on the topology but also on the parameters of gene regulation $K_{j \to i}$. As can be inferred from the earlier studies of FFLs [42], they work at a certain range of parameters. Here, we have found that the evolved network has robustness against parameter variation in $K_{j \to i}$ under fixed network topology. In contrast to perturbation on the topology, the parameter robustness is stronger for long-germ networks than short-germ networks (Fig. 4B; see Figure S12 also for robustness to noisy perturbation in development).

Mutational robustness in evolution could be described by a trade-off between two features of the robustness to network topology and parameters. Comparing the networks evolved under different mutation rate $\mu$ (i.e. the probability of genetic change introduced in a network element per evolutionary generation; see Methods), short germ networks appear more frequently at a higher mutation rate $\mu$ (Fig. 4C). On the other hand, simultaneous expressions of stripes take a shorter developmental time than sequential ones (Fig. 4D).

Hence, long germ modes appear more frequently under a selective pressure for rapid development (Fig. 4E). Transitions between short and long germ-band development occurred during evolution of arthropods [7,8,14-16,49]. This trade-off between the mutational robustness and developmental speed may provide an evolutionary transition from short to long germ mode.

**Future problems.**

Even though we have confirmed correspondence between our models and arthropod in segmentation, there remain some problems that have to be clarified in future: First, peak position of striped pattern in a gene expression is less homogeneous in many of long germ network models (Figs. 1D and S5A) compared with those observed in arthropod. Here, detailed peak position can depend more sensitively on the parameters in development. Even under fixed network topology, the heterogeneity in the peak-to-peak distance in the model was reduced by tuning the parameter values through a suitable selection pressure (Figure S14). Second, we have not so far found any short germ network model with the two roles of gap genes on wild type expression and knockout response described above while they were well documented in *T. castaneum* [64-67]. It might be related to embryo growth around posterior side [15] that was not considered here. Third, the positive FBLs is not a necessary module in our models, while it is necessary to quantitatively reproduce spatial and temporal expression of gap [53] and segment polarity [62] genes in *D. melanogaster*. The present study focuses on rather qualitative aspects of stripe formation and knockout responses to capture a unifying view among diverse striped patterns. The relationship between FFLs and positive FBL will be addressed in evolution of both quantitative and qualitative information in spatial pattern. Last but not the least, evolutionary transition process among the three developmental modes is an important issue to be studied along the line of our study.

**Conclusion**

Our aim here is to elucidate a unifying mechanism behind diverse processes across species. We derive four predictions regarding the network architectures of arthropod segmentation. First, in all long germ arthropods, gene regulatory networks should always exhibit a hierarchical structure composed of multiple FFLs, and the striped pattern of mutants should exhibit a variety of forms. The short germ arthropods, on the other hand, should not necessarily show such a hierarchical structure or a variety in knockout responses. The second is the absolute necessity of a negative FBL for short germ arthropods. Third, an interconnection of FFL and negative FBL is essential for intermediate germ development. And lastly, the double segment periodicity is a signature of spatial organization by serially connected FFLs. For *T. castaneum*, the negative FBL and FFL composed of pair-rule genes [30] should form stripes sequentially and double segmental periodicity, respectively. Although the above predictions should be carefully tested, the overall agreement between our highly abstract model and the well-studied arthropods indicates that the appearance of long, short, and intermediate germ-band development are not by chance but rather by necessity [18,68,69] in the evolution of segmented body plans.

Note added in Proof: In a recent publication [70], evolution of gene network for segmentation is also studied. In particular by focusing on short germ development, they implemented embryo growth at the posterior end to understand ceasing temporal oscillation, known as "clock and wave front" model [71]. They found the mechanism through the interaction of time periodic gene expression and morphogen gradient that moves along with posterior growth. In the present paper, the growth was not concerned and ceasing oscillation rarely appears in short germ mode (Fig. S5B). In contrast, we here have identified for the first time responsible network modules for long and intermediate germ modes as well as short germ mode, and clarified these function. From the analysis of the network architecture, we have explained not only the characteristics of each mode but also many of knockout phenotypes, and predicted arthropod gene network topology.

**Methods**

**Gene network model for development.**

Gene expression is governed by a regulatory network [21,53], in which a single node indicates a single gene, and a connection with an arrow indicates a regulation of a downstream gene #i by an upstream gene #j (see Fig. 2A-C). Architecture of the network is represented by a connection matrix $c_{j \to i}$ where $c_{j \to i}$ = 1, -1 and 0 indicate positive (a red arrow in Fig. 2A-C), negative (a blue arrow), and no regulation, respectively. Expression level of gene #i is represented by the concentration of its product, e.g., protein, $P_i$. The dynamics of the gene expression obeys

$$\frac{\partial P_i}{\partial t} = f\left(P_j; K_{j \to i}\right) - \gamma P_i + D_i \frac{\partial^2 P_i}{\partial x^2} \qquad (1)$$

where $\gamma$ is the degradation rate constant, $D_i$ is the diffusion coefficient of the gene product #i, and $x$ is the position along the anterior-posterior axis in the embryo. The regulation mediated by gene #j follows a Hill equation $f(P_j; K_{j \to i}) = f_+(P_j; K_{j \to i}) \equiv \dfrac{P_j^\alpha}{P_j^\alpha + K_{j \to i}^\alpha}$ for a positive regulation ($c_{j \to i} = 1$) or $f(P_j; K_{j \to i}) = f_-(P_j; K_{j \to i}) \equiv \dfrac{K_{j \to i}^\alpha}{P_j^\alpha + K_{j \to i}^\alpha}$ for a negative regulation ($c_{j \to i} = -1$). Here, $K_{j \to i}$ is a threshold and $\alpha$ is a Hill coefficient. When two genes regulate a gene, combinatorial regulation is introduced (See Supporting Methods S1). For the developmental process, equation 1 is numerically integrated starting from uniform initial concentrations in space for all gene products ($P_i(x) = 0.1$) except for $P_0(x)$. Gene #0 is the maternal factor, which has no regulator. It is synthesized at and diffuses from one pole of the embryo to establish a simple gradient of the form $P_0(x) = A\exp(-x/\lambda)$ at $t = 0$ (See Methods S1). The unit of time $t$ is normalized by the timescale of degradation, $1/\gamma$. Other parameters are: $\alpha = 2$, $\gamma = 1$, $A = 4$, and $\lambda/L = 0.14$ where size of an embryo is given by $L = 100$. 100 cells are arranged in the anterior-posterior direction.

**Evolution of gene network**

A single generation of the evolutionary dynamics is composed of (i) mutation, (ii) development, and (iii) selection (Fig. S1A). (i) From all $N_s$ networks selected at the previous generation, $N_m$ offspring networks are generated by changing the following network elements where $N_s$ and $N_m$ denote the number of selected and offspring networks, respectively: the connection matrix $c_{j \to i}$, the threshold value of each connection $K_{j \to i}$, and the diffusion constant for each product $D_i$ in equation 1. Probability that mutation is introduced in each one of the above elements is defined by the mutation rate $\mu$. The total number of networks in the present generation is $N_s N_m$. (ii) For development, we carried out numerical calculations of equation 1 from $t =0$ to $t =t_{dev}$, and examined the number of stripes in spatial expression pattern of gene #1 for $N_s N_m$ network. (iii) The closeness between the number of stripes for gene #1 at $t =t_{dev}$ and a target number $N_{tar}$ was chosen as a fitness function. Neither detailed position of the stripes, transient behavior of gene #1, nor expression of the other genes is accounted for the fitness. $N_s$ highest networks in the fitness were selected from the $N_s N_m$ networks. These steps complete one generation, and the same procedures are repeated for 2000 generations as a single evolutionary experiment (see evolution of stripe number in Fig. S1B). All elements of the initial networks are set completely at random with no account of prior knowledge of arthropods. We repeated the artificial evolution several hundred times for any given evolutionary condition defined by $N_{tar}$ and $\mu$ ($N_{tar}$ = 10 except for Fig. 4C). For the present work, we choose $N_s$ = 10, $N_m$ = 10, and $t_{dev}$ = 60 except for Figure 4E. (See Methods S1 for further information.)

**Classification of developmental modes.**

When the time required to complete stripes of gene #1 expression is less than a certain threshold $t_{dev}/2$ and all the stripes appear without temporal oscillations in development, the network is classified into a long germ mode. When the time is longer than the threshold and each stripe appears one by one as they oscillate, the network is counted as a short germ mode (See the temporal oscillations in Fig. S9). When a part of the stripes appears within $t_{dev}/2$ and without oscillation, while the remaining stripes appear one by one together with oscillations, the network is classified into an intermediate germ mode. Since the time is different between long and short germ modes (Fig. 4D), the classification is little affected by the choice of the threshold.

**Extraction of core networks.**

We systematically eliminated regulatory connections in the gene networks keeping the number of stripes expressed for gene #1 at $t = t_{dev}$ (See Fig. S4). If the stripes remain unperturbed by the tentative removal of a connection, the connection is eliminated from the network. This process is repeated until no further elimination is possible. The extraction yields a unique network irrespective of the order of elimination for the majority of the networks.

**Network modules.**

When a regulatory connection from a node is looped back to regulate itself via other nodes, it is called a Feed-Back Loop (FBL). A direct auto-regulation is not counted as a FBL. Influence of the feedback regulation in total is classified into negative FBL (see examples marked by Δ in Fig. 2B-C) and positive FBL (e.g., a FBL composed of genes #10 and #13 in Fig. 2B), respectively. When a node regulates another node by two different connections, either directly or indirectly, the sub-network composed of the nodes and their connections are called a Feed-Forward Loop (FFL: e.g., * and + in Fig. 2A-C). The FFL is a loop as structure, but not as a directed network. Here we follow the use of this term by Alon et al., which is widely adopted [58]. Unlike their definition [58], it should be noted that the number of genes within a module is not constrained to three in the present work. This is because it can be analytically shown that the ability of a FFL to form a stripe does not depend on the number of constituent nodes (Fig. S7). We counted the number of FFLs and negative and positive FBLs in each core network (Figs. 2D and S6A-B; See also Fig. S13 for the demonstration of modules extracted from core networks shown in Fig. 2A-C). The number of the core networks used to derive the statistics is 197 for the long germ, 300 for the short germ, and 190 for the intermediate germ networks. When all regulatory pathways from "input" gene #0 to "output" gene #1 pass through a network module (e.g., FFL marked by * in Fig. 2A-C), we defined it as connection in series. When some pathways from gene #0 to #1 go through a module and the others do not (e.g., FFLs marked by + in Fig. 2A and C, and a negative FBL marked by Δ in Fig. 2C), we defined it as connection in parallel.

**Knockout experiments.**

Mutant networks are generated by eliminating a connection or a node from the original network. Elimination of a node is implemented by setting the expression level of the corresponding gene product $P_i$ to 0 throughout development. Likewise, a regulatory connection is eliminated by setting $c_{j \to i}$ to 0. Upon completion of the mutant network development, spatial expression pattern of gene #1 is measured.

**Parameter robustness of striped pattern.**

Maintenance of the stripe number against variation of parameters is investigated. We chose a reference system that exhibits formation of given stripe number, and perturbed the system by randomly modifying its threshold value of gene regulation $K_{j\to i}$ while preserving the network topology. An ensemble of a thousand altered systems was thus generated. Each alternation of the reference system was characterized by the total parameter variation $\delta K$, which was introduced [60] as: $\delta K \equiv \sum_{i,j} \left| \log_e \frac{K'_{j\to i}}{K_{j\to i}} \right|$, where $K_{j\to i}'$ is parameter in the altered system. Development of the altered system was subjected to reaction-diffusion process. Following the developmental process, we measured the fraction of the altered systems that maintain the same stripe number as the original reference system.

**Acknowledgement** We would like to thank S. Sawai, W. W. Chen, T. Gregor, Y. Hiromi, K. Horikawa, A. Kimura, T. Komatsuzaki, T. Mito, S. A. Newman, S. Noji, B. Pfeuty, T. Shibata, K. Sugimura, and T. Takano for carefully reading the manuscript and helpful comments, E. Hoshino, A. Nakajima, and M. Tachikawa for stimulating discussions and help with computing facilities, and C. Fujimoto for help with illustration.

**Author Information** Correspondence should be addressed to K.F (fujimoto@complex.c.u-tokyo.ac.jp)

**Figure Caption**:

**Figure 1.   The evolved networks simulate long, short, and intermediate germ-band development. (A-C)** Schematic representation for the three modes of embryogenesis. **(D-E)** Typical spatio-temporal patterns of the gene #1 during development (upper panel) and snapshots of the final established pattern at $t=60$ (lower panel). The unit $t$ is normalized by the timescale of degradation $1/\gamma$. Ten segmental stripes appear simultaneously at $t\sim10$ in **(D)**, whereas sequentially in **(E)**. In **(F)**, five stripes on the left side first appear simultaneously, and the other five on the right appear sequentially. **(G-I)** Digitized expression at $t=60$ for the genes in the core network (Fig. 2A-C) corresponding to **(D-F)**, respectively (See Figure S2 for quantitative expression pattern and Figure S3 for spatio-temporal development). Gene index is indicated on the right. In **(G)**, expression appears in a gradient (genes #0, #10, #5, and #26), a single stripe (#30), two stripes (#2 and #4), five stripes (#27), and ten stripes (#1) respectively. For genes #11, #10, #13, #17, and #20 in **(H)**, the number of stripes that appear sequentially is about half as many stripes for #1. During short germ-band development, pair-rule genes are also expressed sequentially [10] and show half as many stripes for the segment polarity genes [30]. In **(I)**, spatio-temporal dynamics of genes #21, #25, #4, #10, #15, #8, #31, #3, #17, and #14 agree with expression of gap and pair-rule genes in intermediate germ-band insects [12,13,31]. a.u.; arbitrary unit.

**Figure 2.   FFL, negative FBL and their interconnections characterize the core network architecture. (A-C)** The core networks responsible for generating ten stripes shown in Figure 1D-F, respectively. The number indicates the gene index. **(A)** An example of a core network having no FBL but multiple FFLs; gene #0 to #30, #0 to #6, #26 to #6, #30 to #14, #6 to #14 (connected in parallel marked by +), and #14 to #1 (connected in series marked by *). **(B)** A FBL (marked by a triangle Δ) composed of genes #11, #17, #13, and #10 generates stripes sequentially for the respective genes, whereas a FFL connected in series (marked by *) is composed of #17, #13, #20 and #1 (see Fig. 1I for expression pattern of these genes). **(C)** There exist both FFLs, (indicated by + and *) and a negative FBL (marked by Δ). **(D)** Statistics of core network architectures represented by the fraction of core networks containing a FFL (light green), five FFLs or more (yellow), a negative FBL (pink) and a positive FBL (gray), respectively (See the distribution for number of FFLs and FBLs in Fig. S6A-B). It can be shown from a theory [42] that the minimum number of FFLs required to generate ten stripes is five.

**Figure 3.   Disruption of FFL and negative FBL induce the characteristic defects in arthropods. (A-C)** The knockout index is indicated on the right; e.g., "30→17" and "14" denote removal of a connection between gene #30 and #17 and deletion of gene #14, respectively, whereas "5, 0→5, 5→30" signifies that each response from deleting the gene or connection is the same. The genes and connections belong to either a FFL connected in parallel (colored by yellow green), a FFL connected in series (blue green) or a negative FBL (pink) in the core network (Fig. 2A-C) (see Figure S10 for the spatio-temporal patterns of gene expression). The 1st panel is from the wild type network corresponding to the lower panel in Figure 1D-F. The blue green panels in **(A)** and **(C)** show that stripes are fused in pairs to form a single stripe with absence of every other local minima, except for the lowest panel in **(A)** where a perturbed gene #14 is located at the top of the FFL. **(D)** Genes indicated by the knockout index (see below) agree with arthropod genes in terms of their spatio-temporal development (Figs. S2 and S3) and the patterns of knockout response **(A-C)**. [1] Corresponding to gene #30 and #17 in **A**. [2] #25 in **C**. [3] #27 in **A**, #20 in **B** and #14 in **C**. [4] #14 in **A**. [5] #10, #11, #13 and #17 in **B**. [6] #3 in **C**. *Dm: D. melanogaster, Tc: T. castaneum, Gb: G. bimaculatus, Of: O. fasciatus, Cs: C. salei.*

**Figure 4. Trade-offs between long and short germ modes in development (A, B, D) and evolution (C, E).** **(A)** Frequency distribution of the number of genes is plotted for core networks of long (green) and short (pink) germ modes. **(B)** Robustness of stripe number expressed in gene #1 to parameter variation. By generating an ensemble of systems subjected to parameter change from a given reference system, the fraction of such systems that maintain all the stripes of the original system (upper figure) and the average stripe number among the ensemble (lower figure) are plotted as a function of the total parameter variation $\delta K$ (See Methods). The variations are introduced into the threshold parameters $K_{j \to i}$ in the paths within a core network, while variation into connections without core networks hardly induces stripe defect (inset). **(C)** When the networks are evolved under different mutation rate $\mu$, the ratio between the frequency of long germ mode and that of short germ mode is plotted against $\mu$ (See the absolute frequency in Fig. S11). **(D)** Developmental time required for the stripe formation is shorter for the long germ networks (green) than the short germ networks (pink). For each evolved network, the time it takes to complete formation of the target number of stripes for gene #1 was measured. The distributions in **(A)** and **(D)** and error bar in **(B)** are computed from an ensemble of networks also used to obtain Figure 2D. **(E)** When the networks are evolved under different developmental time constraints, frequencies of appearance of long and short germ modes are plotted against the length of development $t_{dev}$ (see Methods). Target stripe number $N_{tar}$=10 and mutation rate $\mu$=0.05.

**Table 1.   Summary of the three developmental modes in our models.**

| Developmental mode | Long germ | Short germ | Intermediate germ |
|---|---|---|---|
| **Stripe formation** [1] | Simultaneous | Sequential | Combinatorial |
| **Network module** [2] | Multiple FFLs | A negative FBL | FFL and negative FBL |
| **Variety of expression patterns** [3] | Necessary | Not necessary | |
| **Variety of knockout responses** [4] | Necessary | Not necessary | |
| **Mutation rate** [5] | Lower | Higher | |
| **Developmental speed** [6] | Slower | Faster | |

[1] Fig. 1A-F. [2] Fig. 2. [3] Figs. 1G-I, S2 and S7. [4] Fig. 3. [5] Fig. 4C. [6] Fig. 4E.

Long germ mode

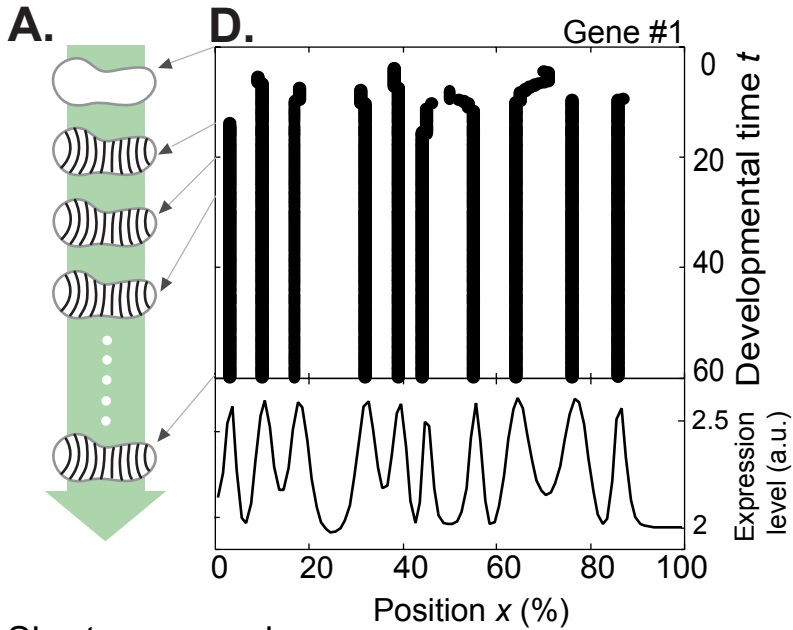

Short germ mode

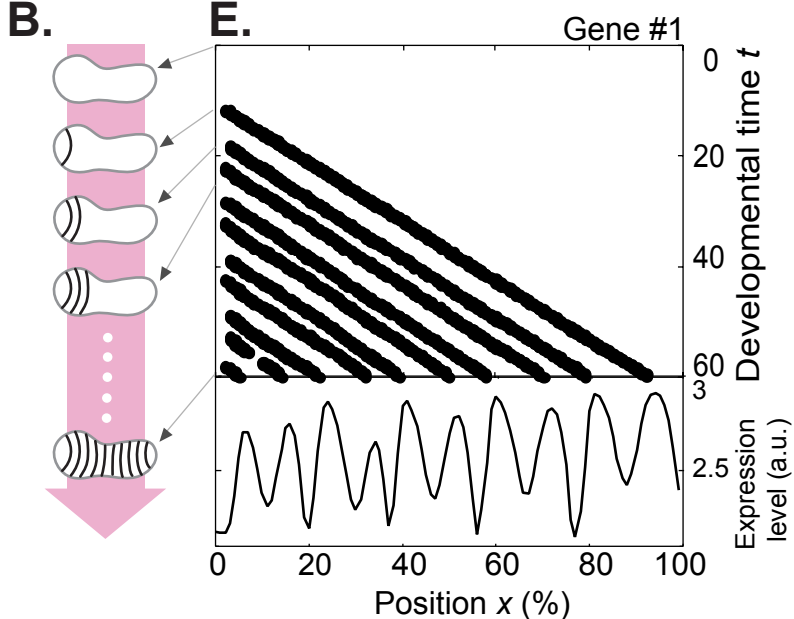

Intermediate germ mode

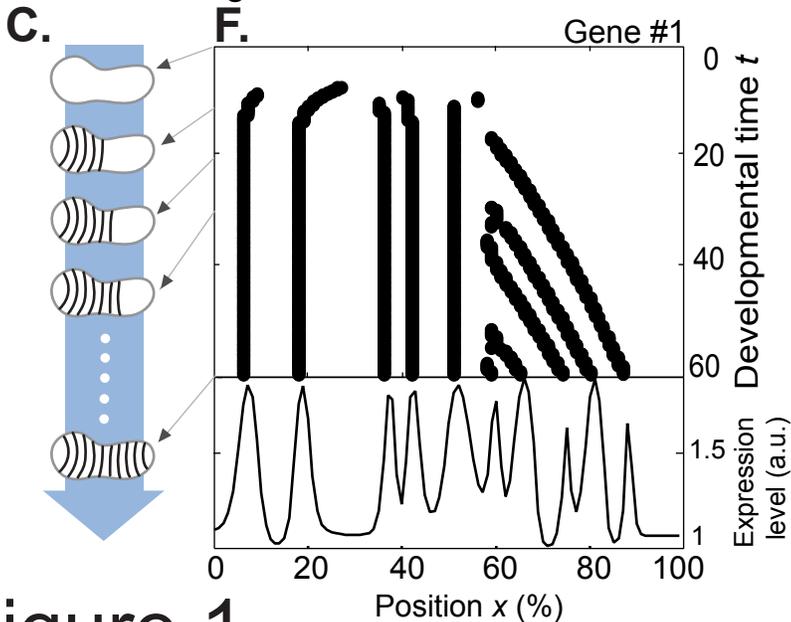

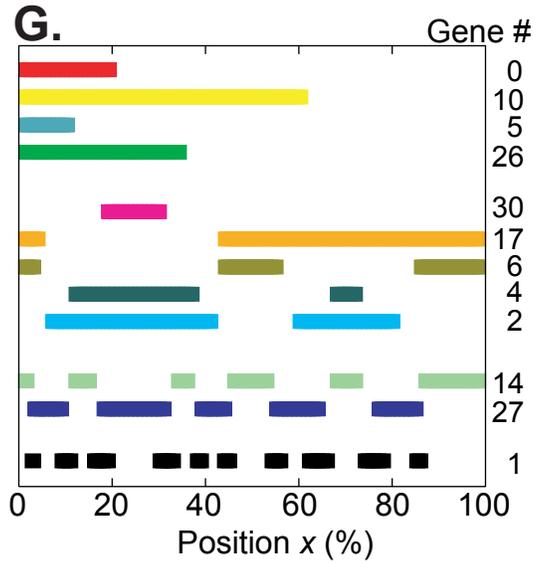

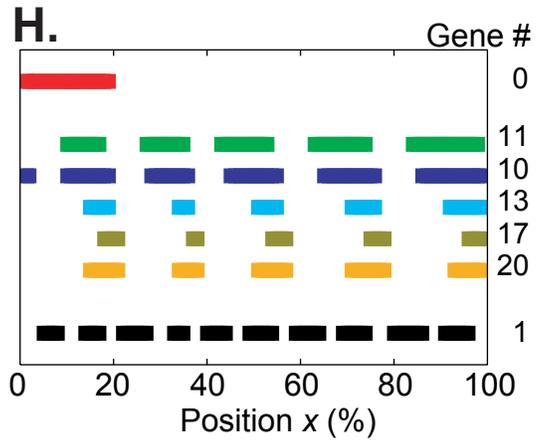

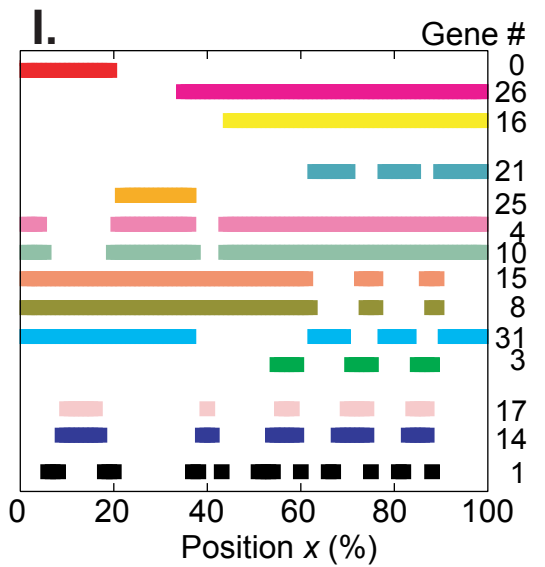

Figure.1

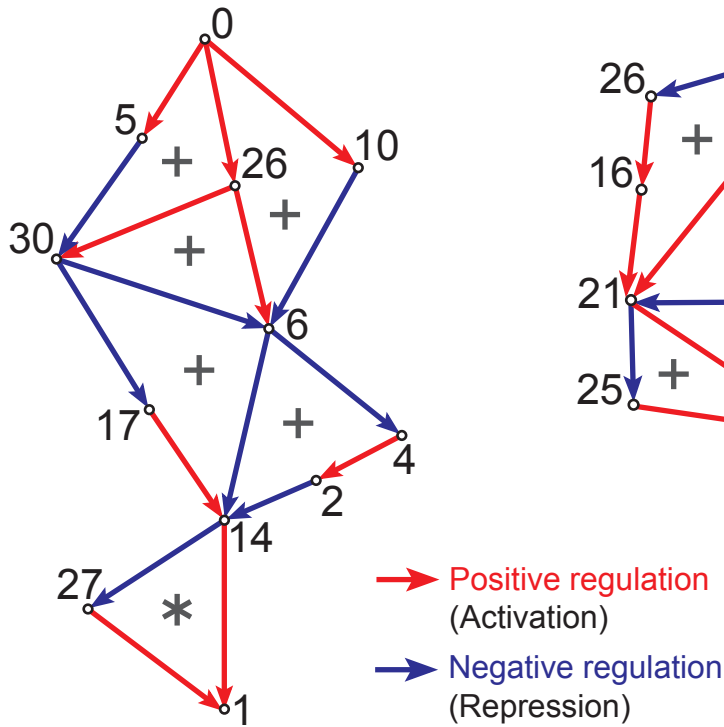
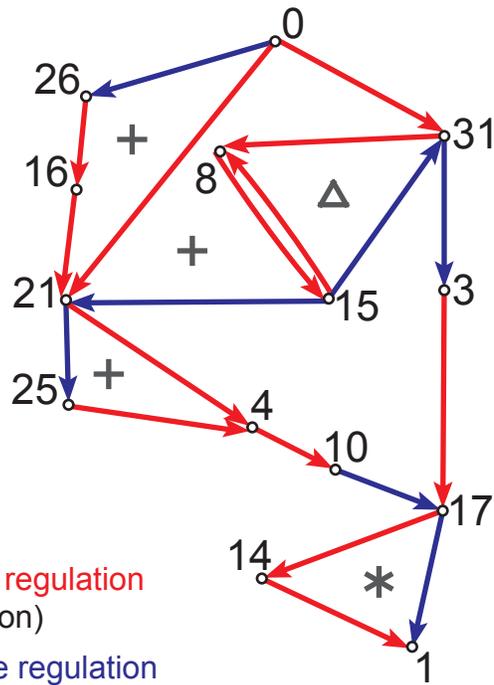
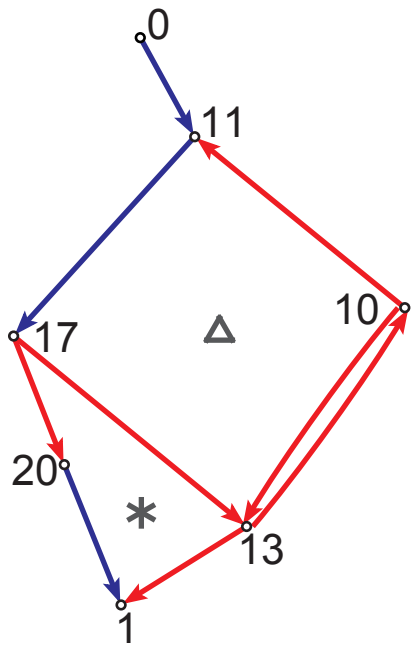
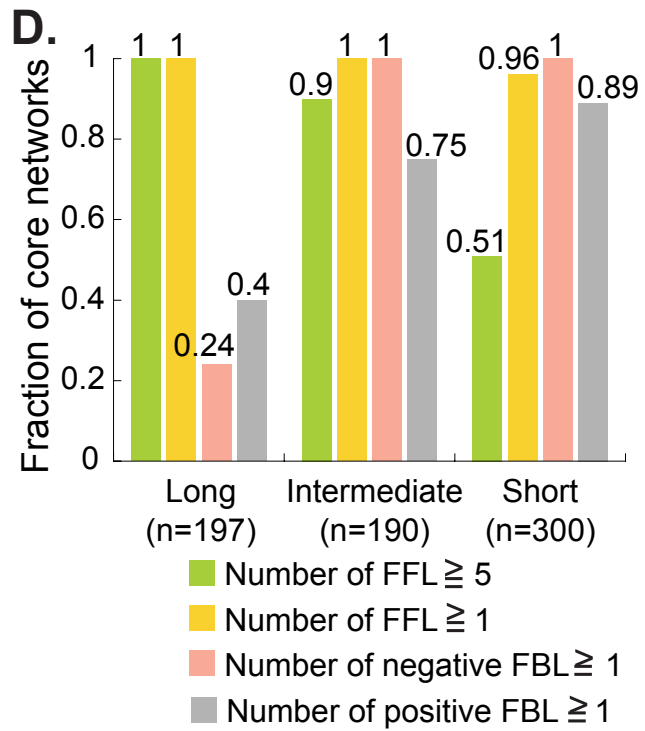

Figure.2

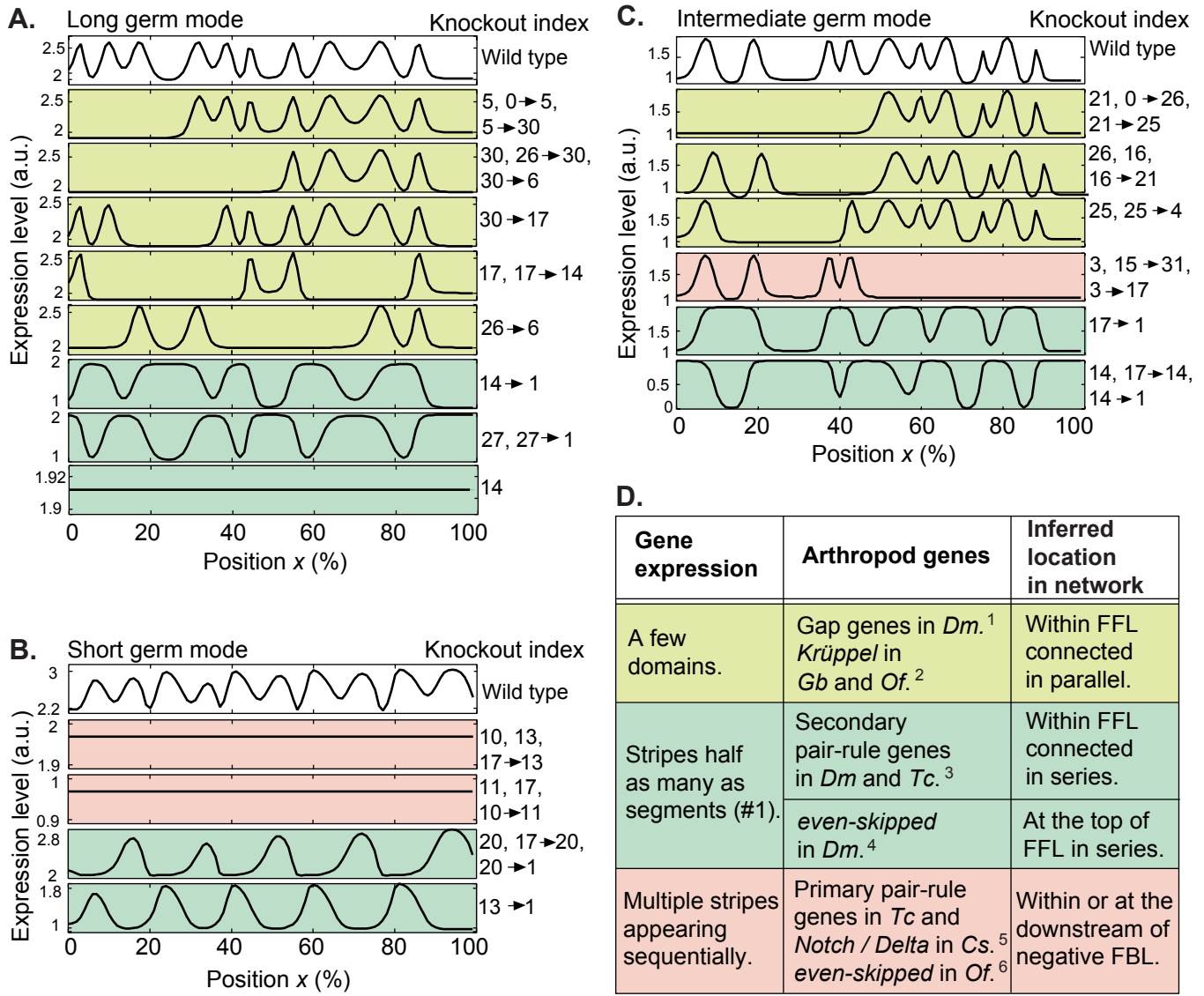

Figure.3

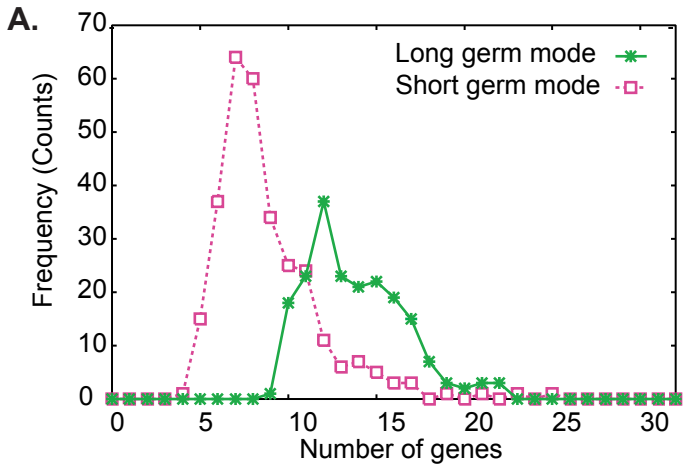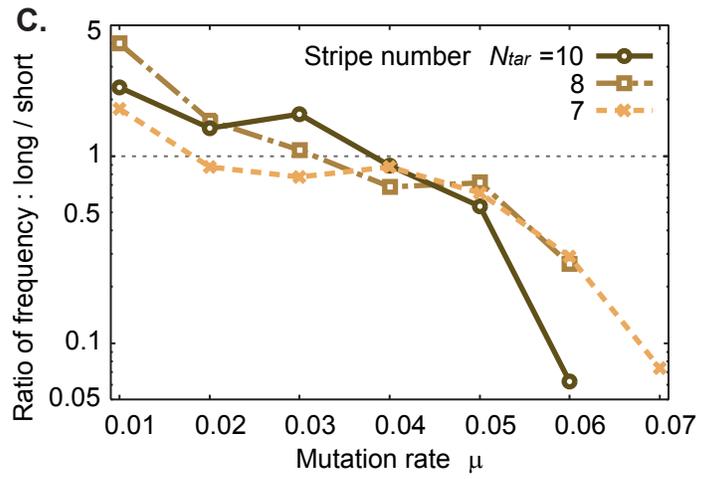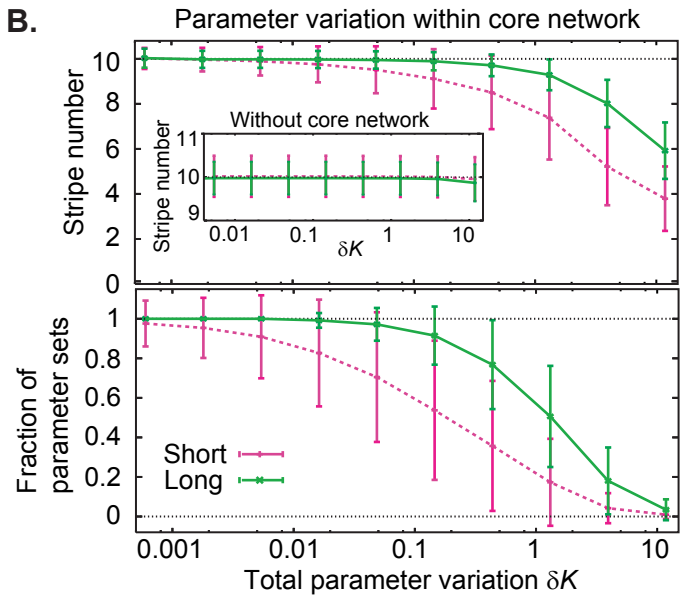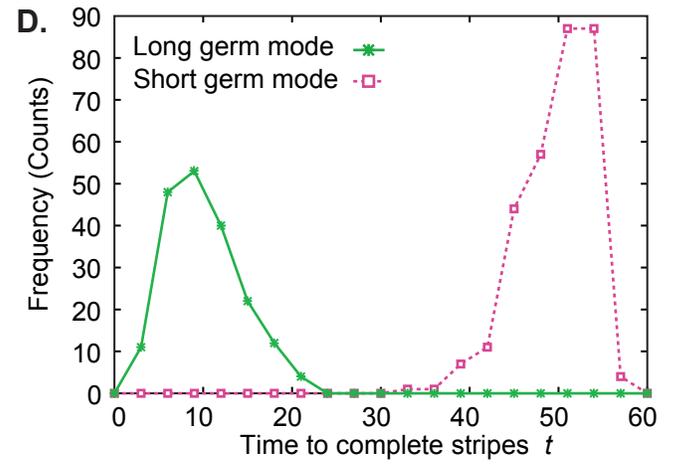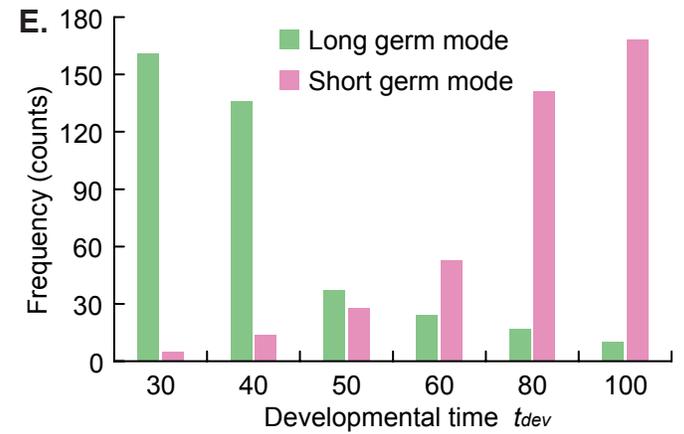

Figure.4